\documentclass{moriond}

\def\Journal#1#2#3#4{{#1} {\bf #2}, #3 (#4)}

\def\NPB{{\em Nucl. Phys.} B}
\def\PLB{{\em Phys. Lett.}  B}

\def\PRD{{\em Phys. Rev.} D}

\def\be{\begin{equation}}
\def\ee{\end{equation}}
\def\bea{\begin{eqnarray}}
\def\eea{\end{eqnarray}}

\newcommand{\Photo}{}

\begin{document}
\vspace*{4cm}
\title{WEAK EQUIVALENCE PRINCIPLE, LORENTZ NONINVARIANCE,  \\AND NUCLEAR DECAYS}

\author{E. FISCHBACH$^{1}$ \footnote{\tt ephraim@purdue.edu}, V. E. BARNES$^{1}$, J. M. HEIM$^{1}$, D. E. KRAUSE$^{2,1}$, J. M. NISTOR$^{1}$}

\address{$^{1}$Department of Physics and Astronomy, Purdue University \\
West Lafayette, IN, 47906 United States \\
$^{2}$Physics Department, Wabash College, Crawfordsville, IN 47933 United States}

\maketitle\abstracts{
We consider three possible manifestations of physics beyond the Standard Model, and the relations among them.  These are  Lorentz non-invariance (LNI), violations of the Weak Equivalence Principle (WEP), and indications of time-varying nuclear decay constants.  We present preliminary results from a new experiment indicating the presence of annual and subannual periodicities in decay data, and discuss their implications for physics beyond the Standard Model.
}

Although the recent discovery of the Higgs boson has reaffirmed our belief in the Standard Model of particle physics, it has also provided a stimulus to search for new physics beyond the Standard Model.  In what follows we discuss possible connections among tests of Lorentz Non-Invariance (LNI), violations of the Weak Equivalence Principle (WEP), and recent evidence for time-varying nuclear  decay constants.  In particular we will suggest that evidence for new physics in any of these areas may also imply new physics in the others.  As we will  note, mounting evidence that nuclear decay rates can be influenced by ambient neutrinos may also be pointing to new physics arising from the preferred frame defined by these neutrinos.

To illustrate the connection between LNI and WEP violation, we consider two models as examples.  In the Nielsen-Picek model \cite{NP} LNI effects are introduced by adding to the usual covariant metric tensor $g_{\mu\nu}(x)$ a constant tensor $\chi_{\mu\nu} = \alpha\, {\rm diag}(1/3,1/3,1/3,-1)$, where $\alpha$ is a constant. [Here we assume $g_{\mu\nu} = \delta_{\mu\nu}$, $x^{\mu} = (\vec{x},x_{4} \equiv ix^{0})$ and $c = 1$.]  For $\alpha \neq 0$ the usual relativistic dispersion relation for a particle of mass $m$ and 4-momentum $p^{\mu} = (\vec{p}, ip^{0})$ ,  
\begin{equation}
-g_{\mu\nu}p^{\mu}p^{\nu} = m^{2},
\end{equation}
becomes
\begin{equation}
-\left(g_{\mu\nu} + \chi_{\mu\nu}\right)p^{\mu}p^{\nu} = m^{2} - \chi_{\mu\nu}p^{\mu}p^{\nu} = m^{2} - \alpha\left(\frac{1}{3}\vec{p}\,^{2} + p_{0}^{2}\right).
\label{g2}
\end{equation}
If we assume that the hypothesized LNI effects in Eq.~(\ref{g2}) arise only in weak interactions through a modification of the W$^{\pm}$ and Z$^{0}$ propagators, then the total inertial mass of a test body $M$ can be written as  \cite{FHTC}
\begin{equation}
M = M_{0} + \alpha B_{w}\left(1 + \frac{4}{3}\vec{v}\,^{2}\right),
\end{equation}
where $M_{0}$ is the total Lorentz invariant portion of the mass, $\vec{v}$ is its velocity, and $\alpha B_{w}$ is the Lorentz non-invariant model-dependent contribution to the mass of each sample. Using energy conservation it is then straightforward to show that the acceleration $a$ of a test mass falling towards the Earth is given by \cite{FHTC}
\begin{equation}
a \simeq \left(1 - \frac{11}{3}\frac{\alpha B_{w}}{M_{0}}\right) g,
\label{a}
\end{equation}
where $g$ is the acceleration due to gravity.
 It follows from Eq.~(\ref{a}) that the difference in acceleration between two test masses \#1 and \#2  is 
\begin{equation}
\frac{\Delta a}{g} = \frac{a_{1} - a_{2}}{g} = -\alpha\frac{11}{3}\left(\frac{B_{w1}}{M_{01}} - \frac{B_{w2}}{M_{02}} \right).
\label{delta a/g}
\end{equation}
This establishes the connection between LNI effects ($\alpha \neq 0$) and WEP violation ($\Delta a/g \neq 0$).

As a second example we consider a modified dispersion relation  \cite{Mattingly} for a particle of mass $m$ and momentum $\vec{p}$,
\begin{equation}
E^{2} = m^{2} + \vec{p}\,^{2} + \frac{\vec{p}\,^{4}}{\mu^{2}},
\label{SME dispersion}
\end{equation}
where $\mu$ is a model-dependent constant.  If the particle is non-relativistic and in the Earth's gravitational field $g$ at height $z$ above the ground level, its energy can be written as
\begin{equation}
E \simeq m + \frac{\vec{p}\,^{2}}{2m} - \frac{\vec{p}\,^{4}}{8m^{3}} + \frac{\vec{p}\,^{4}}{2m\mu^{2}} + mgz.
\label{SME dispersion}
\end{equation}
Then for two different particles \#1 and \#2 falling in the gravitational field, one can show that  the LNI effects arising from the presence of the term proportional to $1/\mu^{2}$ lead to a WEP-violating acceleration difference of the form
\begin{equation}
\frac{\Delta a}{g} = \frac{a_{1} - a_{2}}{g} \simeq 6 \vec{v}\,^{2}\left(\frac{m_{1}^{2}}{\mu_{1}^{2}} - \frac{m_{2}^{2}}{\mu_{2}^{2}} \right),
\label{SME delta a/g}
\end{equation}
where $\vec{v}$ is the particle's velocity.
Hence in this model $\Delta a/g \neq 0$ can arise even when $\mu_{1} = \mu_{2}$ (i.e., the interaction is composition-independent) provided $m_{1} \neq m_{2}$.

Having established the connection between WEP violation and LNI effects, we next ask whether there is any evidence for LNI effects.  Although there is no direct evidence at present, there is both direct and indirect evidence that ambient solar and cosmic neutrinos are in fact interacting with our detection systems.  Since cosmic neutrinos (i.e., relic big-bang neutrinos) define a preferred coordinate frame with respect to which the Earth is moving, LNI effects could in principle arise if these neutrinos interact with local experiments.  The same is also true for solar neutrinos which have been detected in terrestrial experiments.\cite{Yoo}  Here we focus on the possibility that resent observations of time-dependent nuclear decay parameters could also arise from interactions between background neutrinos and terrestrial detectors.  If so, these effects could represent evidence for LNI contributions, and by extension, WEP violations as well.

In Table~1 of Ref.~\cite{O'Keefe} a summary is presented of earlier results indicating time-varying nuclear decay rates.  Although the most common periodic signals seen in those data are annual variations, the most significant are those associated with solar rotation \cite{Solar rotations},  and with solar storms \cite{Solar storms}, since these cannot reasonably be attributed to seasonal variations in the efficiencies of the detectors in the respective experiments.  Further support for the inference that the time varying effects are not simply variations in detector efficiencies comes from experiments in which dissimilar variations were seen in the decays of different isotopes being recorded by the same detectors.\cite{BNL,Ellis,Parkhomov}

Here we present preliminary results from a repetition by our group of the original BNL experiment \cite{BNL} which measured the half-life of $^{32}$Si using $^{36}$Cl as a comparison standard.  Our experiment utilized both the same samples and the same sample-changing system as in the original experiment, but included an updated detector and electronics.  As in the original experiment, data were taken in alternating half-hour runs on the $^{32}$Si and $^{36}$Cl samples.    This insured that the same long term variations in detector efficiencies would be present in the daily count rates of each isotope, and would thus cancel when the $^{32}$Si/$^{36}$Cl ratio was determined daily.   In contrast to the original experiment, which acquired data for only a few days each month for a period of 4 years, our experiment has run continuously for a period of 2 years in an environment where the influences of variations in temperature, pressure, humidity, and magnetic fields have been controlled and monitored.

\begin{figure}[t]
\centerline{\includegraphics[height=7cm]{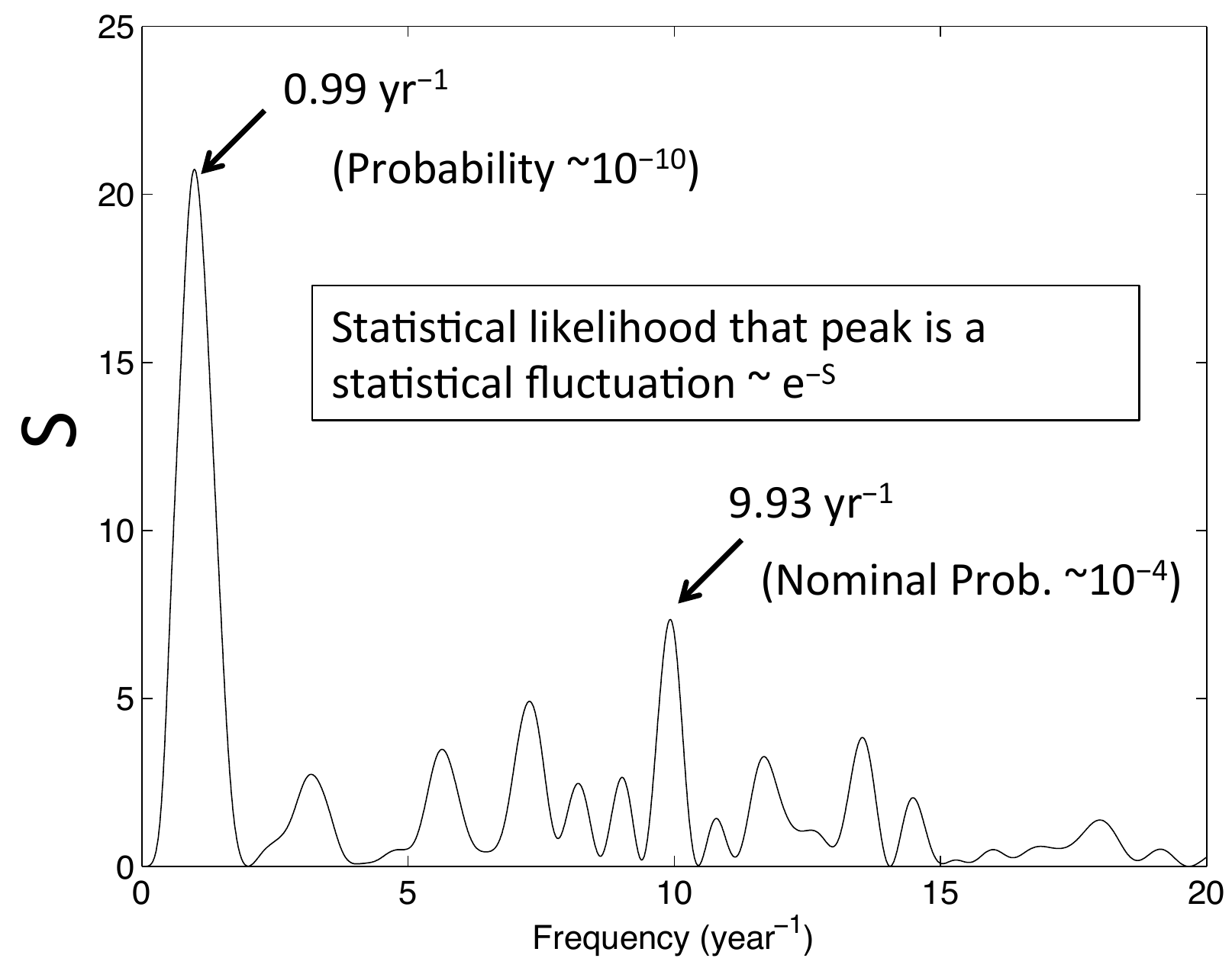}}
\caption[]{Power spectrum analysis of time series of  $^{32}$Si/$^{36}$Cl data showing evidence of 1-year and 0.1-year periods.  As noted in the text, the latter cannot be attributed to any known environmental effect on our detection system, and has been observed in other decay experiments.}
\label{power series figure}
\end{figure}

Figure~\ref{power series figure} presents a power spectrum analysis of the time series of data formed by taking the daily ratios of the $^{32}$Si/$^{36}$Cl data.  
We see immediately a strong annual signal along with a signal with a frequency of $\sim10$/year similar to the signals found  in other decay experiments.\cite{Sturrock Park}  These signals cannot be attributed to any known environmental effect on our detection system.

\begin{figure}[t]
\centerline{\includegraphics[width=13cm]{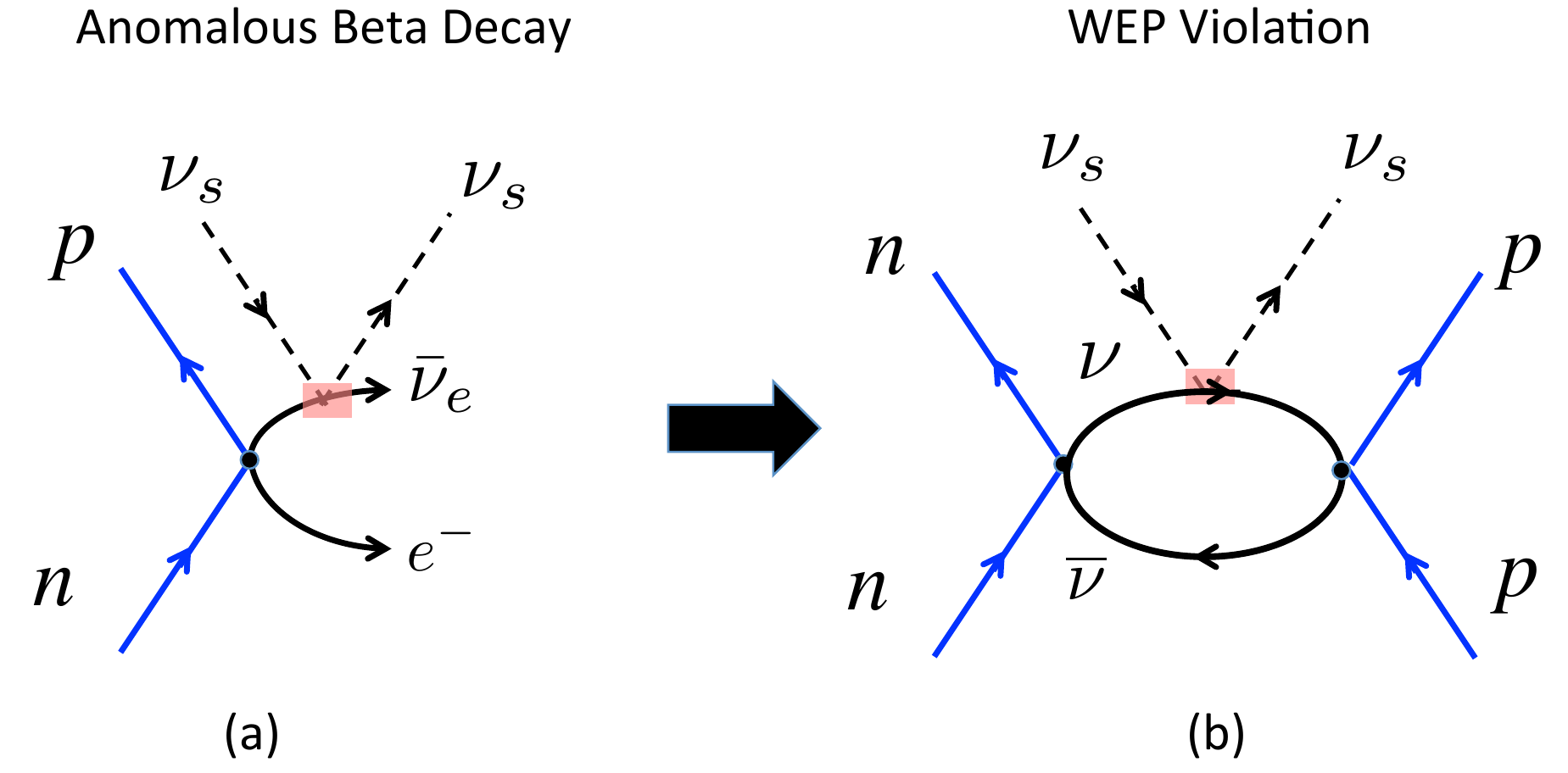}}
\caption[]{(a) Feynman diagram of an anomalous beta decay of a neutron, where the emitted electron anti-neutrino ($\overline{\nu}_{e}$) interacts with a solar neutrino ($\nu_{s}$).  (b) The corresponding diagram of a solar neutrino interacting with a virtual neutrino in a nucleus, leading to a violation of the WEP.}
\label{Decay WEP Figure}
\end{figure}

A possible explanation for these decay anomalies suggested by the observed frequencies and correlations with solar storms could be an interaction involving solar or relic neutrinos.\cite{Space Sci}  A possible interaction shown in Fig.~\ref{Decay WEP Figure}(a) involves a solar neutrino scattering off the  electron anti-neutrino emitted in a  beta decay, which would lead to a modification of the decay rate.  If such a process existed, then there would also be a similar scattering off a virtual neutrino found in the 2-neutrino-exchange interaction between nucleons as shown in Fig.~\ref{Decay WEP Figure}(b).  Such an interaction would lead to a composition-dependent force and apparent violations of the WEP in gravity experiments.\cite{2 neutrino}  Such WEP violations could show up in the forthcoming space-based MICROSCOPE experiment, scheduled to be launched in 2016.\cite{Touboul}

\begin{figure}[t]
\centerline{\includegraphics[width=14cm]{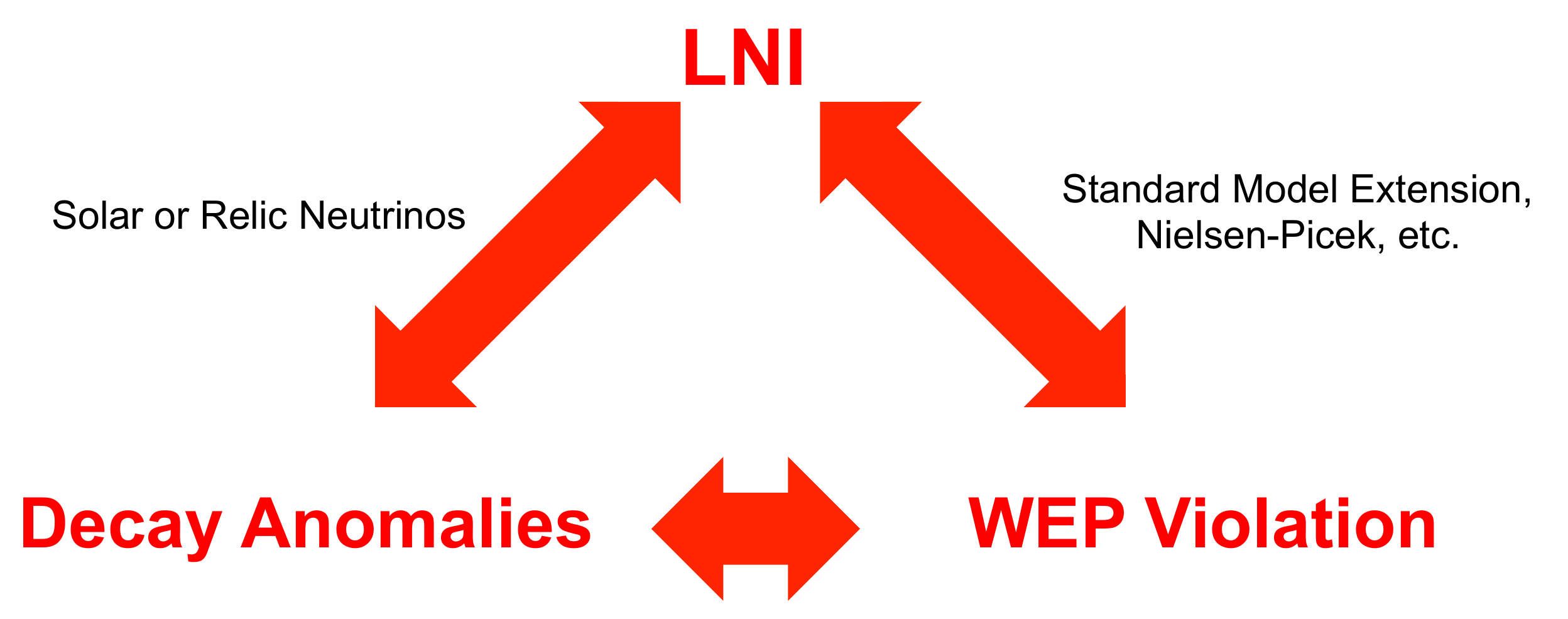}}
\caption[]{The relationships among Lorentz non-invariance, WEP violation, and decay anomalies.}
\label{Fig3}
\end{figure}

Figure~\ref{Fig3} summarizes the principal conclusions of this paper.  Lorentz non-invariance (LNI) almost inevitably leads to violations of the Weak Equivalence Principle (WEP).  The  background of solar and relic neutrinos yields  preferred directions in space that could produce interactions leading to apparent LNI, and to variations in nuclear decay rates.  Such interactions would inevitably produce composition-dependent interactions that could appear in experiments testing the WEP.  Hence, anomalies observed in any one of these three areas has consequences for  the others.

\section*{Acknowledgments}

The authors wish to thank J. Jenkins, D. Koltick, T. Mohsinally, J. Scargle, and G. Steinitz for helpful discussions, and to P. A. Sturrock for the spectral analysis leading to Figure~1.  We are deeply indebted to D.E. Alburger, and G. Harbottle for making their samples and detection systems available to us.

\end{document}